\documentclass[a4paper]{IEEEtran}
\usepackage{cite}
\ifCLASSINFOpdf
  \usepackage[pdftex]{graphicx}
\else
  \usepackage[dvips]{graphicx}
\fi
\usepackage[T1]{fontenc}
\usepackage[cmex10]{amsmath}
\usepackage{array}
\usepackage{theorem}
\usepackage[caption=false,font=footnotesize]{subfig}
\usepackage{fixltx2e}
\usepackage{url}
\usepackage{acronym}
\usepackage{comment}
\usepackage{soul,color}
\usepackage{multirow}
\usepackage{amsmath}
\usepackage{amssymb}

{\theorembodyfont{\rmfamily}
   }
{\theorembodyfont{\rmfamily}
   }
{\theorembodyfont{\rmfamily}
   }
{\theorembodyfont{\rmfamily}
   }

{\theorembodyfont{\rmfamily}
   }
{\theorembodyfont{\rmfamily}
   }

\acrodef{LDPC}{low-density parity-check}
\acrodef{MDPC}{moderate-density parity-check}
\acrodef{QC}{quasi-cyclic}
\acrodef{QC-LDPC}{quasi-cyclic low-density parity-check}
\acrodef{QC-MDPC}{quasi-cyclic moderate-density parity-check}
\acrodef{RSA}{Rivest, Shamir, Adleman}
\acrodef{BF}{bit flipping}
\acrodef{SPA}{sum product algorithm}
\acrodef{RDF}{random difference families}
\acrodef{ISDA}{information set decoding attacks}
\acrodef{DCA}{dual code attacks}
\acrodef{WF}{work factor}
\acrodef{BER}{bit error rate}
\acrodef{FER}{frame error rate}
\acrodef{CER}{codeword error rate}
\acrodef{BLER}{block error rate}
\acrodef{PLS}{Physical layer security}
\acrodef{SNR}{signal-to-noise ratio}
\acrodefplural{SNR}[SNRs]{signal-to-noise ratios}
\acrodef{eBCH}{extended Bose-Chaudhuri-Hocquenghem}
\acrodef{CCs}{convolutional codes}
\acrodef{CC}{convolutional code}
\acrodef{UB}{union bound}
\acrodef{TUB}{truncated union bound}
\acrodef{QSFC}{quasi-static fading channel}
\acrodefplural{QSFC}[QSFCs]{quasi-static fading channels}
\acrodef{FFC}{fast fading channel}
\acrodef{CSI}{channel state information}
\acrodef{AWGN}{additive white Gaussian noise}
\acrodef{NMS}{normalized min-sum}
\acrodef{LLR}{log likelihood ratio}
\acrodef{LLR-SPA}{log-likelihood ratio sum-product algorithm}
\acrodef{BCC}{broadcast channel with confidential messages}
\acrodef{UEP}{unequal error protection}
\acrodef{BPSK}{binary phase shift keying}
\acrodef{QAM}{quadrature amplitude modulation}
\acrodef{ML}{maximum likelihood}
\acrodef{BEC}{binary erasure channel}
\acrodef{BSC}{binary symmetric channel}
\acrodef{AONT}{all-or-nothing transform}
\acrodefplural{AONT}[AONTs]{all-or-nothing transforms}
\acrodef{PC}{protection class}
\acrodefplural{PC}[PCs]{protection classes}
\acrodef{MIMO}{multiple-input multiple-output}
\acrodef{OSD}{ordered statistics decoding}
\acrodef{ISD}{information set decoding}
\acrodef{DMC}{discrete memoryless channel}
\acrodefplural{DMC}[DMCs]{discrete memoryless channels}
\acrodef{MAP}{maximum a posteriori}
\acrodef{PEG}{Progressive Edge Growth}

\begin{document}

\title{Performance assessment and design of finite length LDPC codes for the Gaussian wiretap channel 
\thanks{This work was supported in part by the MIUR project ``ESCAPADE''
(Grant RBFR105NLC) under the ``FIRB -- Futuro in Ricerca 2010'' funding program.}
}
\author{\IEEEauthorblockN{Marco Baldi, Giacomo Ricciutelli, Nicola Maturo, Franco Chiaraluce,\\}
\IEEEauthorblockA{DII, Universit\`a Politecnica delle Marche,\\
Ancona, Italy\\
Email: \{m.baldi, n.maturo, f.chiaraluce\}@univpm.it}, g.ricciutelli@pm.univpm.it }

\maketitle

\pagestyle{empty}
\thispagestyle{empty}

\begin{abstract}
In this work we study the reliability and secrecy performance achievable by practical
\ac{LDPC} codes over the Gaussian wiretap channel.
While several works have already addressed this problem in asymptotic conditions, i.e., under
the hypothesis of codewords of infinite length, only a few approaches exist for the finite length regime.
We propose an approach to measure the performance of practical codes and compare it with that 
achievable in asymptotic conditions.
Moreover, based on the secrecy metrics we adopt to achieve this target, we propose a code optimization algorithm
which allows to design irregular \ac{LDPC} codes able to approach the ultimate 
performance limits even at moderately small codeword lengths (in the order of $10000$ bits).
\end{abstract}


\section{Introduction}
\label{sec:Intro}

Coding for the Gaussian wiretap channel is a well-established research topic, but there are some partially unsolved and challenging problems.
One of these problems is to study the secrecy performance in the finite code length regime, and to design optimized finite length codes.
One of the most common metrics to assess the performance of finite length codes used for transmissions is the average \ac{BER} 
achieved by using some (possibly optimal) decoder.
On the other hand, the secrecy performance over wiretap channels is classically measured using information-theoretic metrics, like
the secrecy capacity, and in asymptotic conditions (e.g., infinite code length and random coding).

For example, in \cite{Thangaraj2007b}, the authors consider a \ac{DMC} (that is, a \ac{BEC} or \ac{BSC}) model for both the main and wiretapper's channels,
and design optimized regular \ac{LDPC} codes for these channels.
They show that their approach achieves the secrecy capacity when the wiretap channel consists of symmetric \acp{DMC}. 
No continuous channels are considered, and the secrecy capacity is achieved in the asymptotic regime (i.e., with infinite length codes).

The \ac{BER} as a secrecy metric has instead been used in \cite{Klinc2011b}, where a coding scheme able to achieve a \ac{BER} very close to $0.5$ for the eavesdropper and very low for the authorized channel is proposed.
In \cite{Klinc2011b}, the authors use differential evolution to design optimized \ac{LDPC} codes able to achieve the desired \ac{BER} targets while keeping the quality ratio between the main and the eavesdropper's channels (named security gap) as small as possible.
The proposed coding scheme is based on puncturing and, thanks to the \ac{BER}-based analysis, is applicable at finite block lengths.
A similar solution, but without the need of puncturing, has been proposed in \cite{Baldi2012}, and extended in \cite{Baldi2014} to the case of parallel channels.

A bridge between information theoretic and error rate-based secrecy measures is presented in \cite{WongWong2011}, where however the main goal is to propose a secret key sharing scheme for the wiretap channel, and the presence of an error-free public channel between the source and destination is considered, which helps the secret sharing process. 
By using regular \ac{LDPC} codes, the authors show that the key capacity can be achieved in the asymptotic regime.
Irregular \ac{LDPC} codes are instead considered for the finite code length regime, and a density evolution based linear program is used to design them.
The same approach is followed in \cite{WongWong2011a} to assess the performance of punctured \ac{LDPC} codes over the Gaussian wiretap channel.

Inspired by such works, in this paper we study the performance of finite length \ac{LDPC} codes over the Gaussian wiretap channel, by defining suitable metrics to assess how far they are from optimality, which is achieved in asymptotic conditions.
This permits us to explore the capacity-equivocation regions of these codes in the finite length regime, and without using puncturing.
We also propose a twofold code optimization tool which allows to design optimal codes in terms of the considered metrics.
Similar twofold code optimizations have been proposed for the relay channel \cite{Chakrabarti2007b, Wang2013, Khattak2014}, but no solution has been presented for the wiretap channel, at our best knowledge.
We show that our approach allows to achieve great flexibility in the choice of the system parameters, as well as higher security levels with respect to previous solutions based on punctured \ac{LDPC} codes \cite{WongWong2011a}.

The organization of the paper is as follows. 
In Section \ref{sec:Model} we present the system model and the metrics we use to assess performance in asymptotic and finite length conditions. 
In Section \ref{sec:CodeDesign} we describe the code design requirements.
In Section \ref{sec:CodeOptimization} we propose our code optimization approach. 
In Section \ref{sec:NumericalResults} we provide and discuss some numerical results. 
Finally, Section \ref{sec:Conclusion} concludes the paper.

\section{System model and metrics}
\label{sec:Model}

We consider the classical Gaussian wiretap channel model, in which a sender, named Alice, transmits a secret message $\mathcal M$.
She encodes her message into the $n$-symbol codeword $X^n$, which uniquely depends on $\mathcal M$ and on some random message $\mathcal R$ generated by Alice.
We consider binary coding, therefore $X^n$ actually is an $n$-bit codeword.
If the secret message is $k_s$ bits long and the random message is $k_r$ bits long, the code rate is $R_c = (k_s + k_r)/n = k/n$.
The secret message rate, instead, is $R_s = k_s/n$.

Transmission occurs over a Gaussian channel for both the authorized receiver, named Bob, and the eavesdropper, named Eve.
The noisy codewords received by Bob and Eve are denoted by $Y^n$ and $Z^n$, respectively.
In order to achieve successful transmission of $\mathcal M$ over this channel, both the following targets must be fulfilled:
\begin{enumerate}
\item $\mathcal M$ must be reliably decoded by Bob, i.e., with a sufficiently small error rate (\textit{reliability target}),
\item the information about $\mathcal M$ gathered by Eve must be sufficiently small (\textit{security target}).
\end{enumerate}

Concerning the reliability target, in ideal conditions (i.e., infinite code length and random coding) the channel capacity
can be used as the ultimate code rate limit.
In the finite length regime, instead, a practical code must be designed to allow Bob to achieve a sufficiently low error rate
in decoding the secret message.
Concerning the security target, some classical information theoretic secrecy metrics are only useful in the 
asymptotic regime. In fact, denoting by ${\rm I}(x; y)$ the mutual information between $x$ and $y$, we have \cite{Bloch2011}:
\begin{itemize}
\item Strong secrecy when the total amount of information leaked about $\mathcal M$ through observing $Z^n$ goes to zero as $n$ goes to infinity, i.e., $\displaystyle \lim_{n \rightarrow \infty}{\rm I}(\mathcal M; Z^n) = 0$.
\item Weak secrecy when the rate of information leaked about $\mathcal M$ through observing $Z^n$ goes to zero as $n$ goes to infinity, i.e., $\displaystyle \lim_{n \rightarrow \infty}{\rm I}(\mathcal M; Z^n)/n = 0$.
\end{itemize}
So, these metrics are not useful in order to assess the performance in finite length conditions and compare it with that in the asymptotic regime.

However, another metric can be exploited, which was already used in Wyner's original work \cite{Wyner1975}.
According to \cite{Wyner1975}, transmission is accomplished in perfect secrecy when the wiretapper equivocation rate on the secret message, $R_e = \frac{1}{n} {\rm H}(\mathcal  M|Z^n)$, with ${\rm H}(\cdot)$ denoting the entropy function, equals the entropy of the data source.
We consider independent and identically distributed secret messages, therefore the source entropy rate is equal to $R_s$.
So, perfect secrecy is achieved when the equivocation rate $R_e$ equals the secret message rate $R_s$, i.e.,
\begin{equation}
\widetilde{R_e} = R_e/R_s = 1.
\label{eq:PerfectSecrecy}
\end{equation}
$\widetilde{R_e}$ is called fractional equivocation rate.

Actually, the ultimate limit achievable by the equivocation rate is the secrecy capacity $C_s = C_B - C_E$, where $C_B$ and $C_E$ are Bob's and Eve's channel capacities, respectively. 
For a binary-input channel with \ac{AWGN} and \ac{SNR} $\gamma$, the capacity is given by the following expression:
\begin{equation}
\label{eq:capacity}
C\left( \gamma \right) = 1 - \frac{1}{\sqrt{2\pi}} \int_{-\infty}^{\infty} e^{-\frac{\left (y - \sqrt{\gamma} \right )^2}{2}} \log_2 \left(1 + e^{-2y\sqrt{\gamma}} \right) dy.
\end{equation}
Then, the target is to maximize $R_e$ in such a way as to approach the secrecy capacity. On the other hand, when considering finite length codes, it is expected that $C_s < R_s$ and another valuable issue is the evaluation of the gap between the secret message rate and the secrecy capacity. Numerical examples will be presented in Section \ref{sec:NumericalResults}.

\begin{figure}[!t]
\begin{centering}
\includegraphics[width=90mm,keepaspectratio]{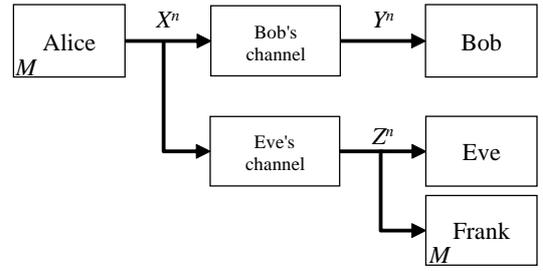}
\caption{Wiretap channel model employed in the study.}
\label{fig:Wiretap}
\par\end{centering}
\end{figure}

Concerning the computation of the equivocation rate, it can be shown that \cite{WongWong2011a}:
\begin{equation}
\begin{split}
R_e & = \frac{1}{n} \left[{\rm H}(X^n) - {\rm I}(X^n; Z^n)  + {\rm H}({\mathcal M}|Z^n, X^n) \right.\\
& \left. - {\rm H}(X^n|{\mathcal M}, Z^n)\right].
\label{eq:Re}
\end{split}
\end{equation}
From \eqref{eq:Re} it results that this formulation of Eve's equivocation rate requires to compute the quantity ${\rm H}(X^n|{\mathcal M}, Z^n)$,
that is, the entropy of $X^n$ conditioned to receiving $Z^n$ and knowing the secret message ${\mathcal M}$.
Eve obviously does not know the secret message, therefore we suppose the existence of another (fictitious) receiver in the same position as Eve's, knowing the secret message ${\mathcal M}$.
We denote such a receiver as Frank: he receives the same vector $Z^n$ as Eve but, differently from Eve, he has perfect knowledge of the secret message ${\mathcal M}$.
Then, he tries to decode $Z^n$ for recovering the random message ${\mathcal R}$, which is the only source of uncertainty for Frank in order to reconstruct $X^n$.
The resulting wiretap channel model is schematically depicted in Fig. \ref{fig:Wiretap}. The letter $M$ inside Alice's and Frank's boxes points out that the message is known to both Alice and Frank.

Let us suppose that, in these conditions, Frank experiences a decoding error probability (or \ac{CER}) equal to $\eta$.
By Fano inequality we have ${\rm H}(X^n|{\mathcal M}, Z^n) \le 1 + k_r \eta$.
We also have ${\rm H}(X^n) = k$ and ${\rm H}({\mathcal M}|Z^n, X^n) \leq {\rm H}({\mathcal M}|X^n) = 0$.
Concerning Eve's channel mutual information ${\rm I}(X^n; Z^n)$, we could obtain a tight upper bound on it as proposed in \cite{Polyanskiy2010},
by taking into account the code length and the target error rate.
However, by using the classical bound ${\rm I}(X^n; Z^n) \leq n C_E$, we obtain a limit value which is independent
of Eve's error rate. Such a value cannot be overcome even if Eve's error rate changes, therefore it represents a conservative
choice for our purposes.
Based on these considerations, we can find a lower bound on Eve's equivocation rate about the secret message as \cite{WongWong2011a}:
\begin{equation}
\begin{split}
R_e & \ge \frac{1}{n} \left[k - n C_E  - k_r \eta - 1 \right] = \\
& = R_c - C_E - (R_c - R_s)\eta - \frac{1}{n} = R_e^*.
\label{eq:ReBound}
\end{split}
\end{equation}

By looking at \eqref{eq:ReBound}, it is evident that this metric is well suited to assess the secrecy performance of practical, finite length codes.
In fact, the code length is taken into account, and the error rate experienced by Frank can be estimated for practical codes through numerical simulations.
Its value obviously depends on Frank's \ac{SNR}, which is the same as Eve's, and therefore, according to \eqref{eq:capacity}, it determines $C_E$.
It follows that, for a fixed code length and rate, the equivocation rate can be maximized by optimizing the choice of the pair $(\eta, C_E)$.

\section{Code design}
\label{sec:CodeDesign}

An \ac{LDPC} code with rate $R_c=k/n$ is defined through its parity-check matrix $\mathbf H$ of size $(n-k) \times n$.
Alternatively, the \ac{LDPC} code can be represented through a Tanner graph, that is a bipartite graph composed of variable and check nodes, which correspond to the codeword bits and the parity-check equations, respectively.
Noting by $h_{ij}$ the $(i, j)$-th element of $\mathbf H$, there is an edge between the $j$-th variable node and the $i$-th check node iff $h_{ij} = 1$.
The number of edges connected to a node is called degree of that node.
The following two polynomials are commonly used to denote the variable and check node degree distributions:

\begin{equation}
\lambda(x)=\sum_{i=2}^{d_{v}}{\lambda_i x^{i-1}}, \quad \quad \rho(x)=\sum_{j=2}^{d_{c}}{\rho_j x^{j-1}}
\label{DegreeDistribution}
\end{equation} 
where $d_{v}$ and $d_{c}$  are the maximum variable and check node degrees, respectively. 
In $\lambda\left(x\right)$ $\left(\rho\left(x\right)\right)$, the coefficient $\lambda_i$ $\left(\rho_j\right)$ coincides with the fraction of edges connected to the variable $\left(\text{check}\right)$ nodes having degree $i$ $\left(j\right)$. 
Therefore, $\lambda(x)$ and $\rho(x)$ are defined from the edge perspective.
The code rate can be expressed as:

\begin{equation}
R_c=1-\frac{\sum_{i=2}^{d_{v}}\rho_i/i}{\sum_{j=2}^{d_{c}}\lambda_j/j}.
\label{Rate}
\end{equation} 

The most common \ac{LDPC} code decoding algorithm, which is an instance of the well-known belief propagation principle, is based on the exchange of soft messages about each received bit between the nodes of its Tanner graph.
Therefore, the performance of an \ac{LDPC} code depends on the connections among the nodes of its Tanner graph.
Indeed, a variable node with a greater number of connected edges has more parity-check equations which verify its associated bit. 
On the other hand, check nodes with low degrees correspond to parity-check equations with less unknowns.
The optimization of the code performance under message passing decoding consists in finding the best tradeoff between these two effects, and this usually requires irregular degree distributions.
The well-known density evolution algorithm, proposed in \cite{Chung 2001}, aims at optimizing the pair $\left(\lambda(x), \rho(x)\right)$ based on the statistics of the decoder messages.
However, differently from classical transmission problems, in our setting the same code (chosen by Alice) is used by three receivers: Bob, Eve and Frank, and the code optimization
should take this into account.

\begin{figure}[!t]
\begin{centering}
\includegraphics[width=80mm,keepaspectratio]{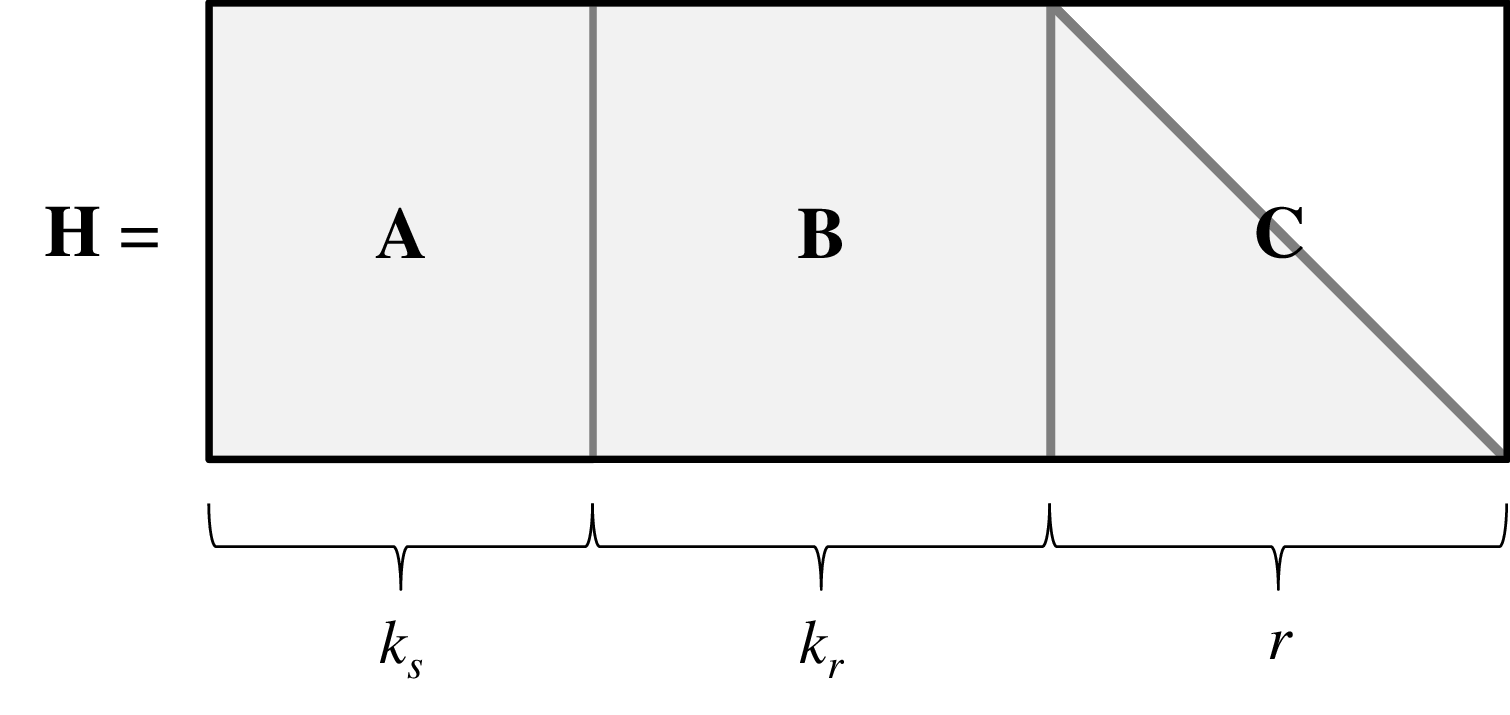}
\caption{Parity-check matrix of the considered codes.}
\label{fig:matrixH}
\par\end{centering}
\end{figure}

Let us consider a systematic encoder and let us suppose that the transmitted codeword is $\mathbf c = [\mathcal M | \mathcal R | \mathcal P]$, where $\mathcal M$ is the $k_s$-bit secret message,
$\mathcal R$ is the $k_r$-bit random message and $\mathcal P$ is the $r$-bit redundancy vector added by the encoder.
Obviously, systematic encoding shall be avoided in security applications, especially if source coding is not optimal.
In fact, in such a case, Eve could look at the systematic part of the received codeword and gather some information about the secret message parts which are less affected by errors.
In practical systems, systematic encoding can be easily avoided by scrambling the information bits prior to encoding \cite{Baldi2012}.
Having this clearly in mind, for our code design and analysis purposes it is convenient to keep the assumption of systematic encoding. 
Under this hypothesis, the code parity-check matrix can be divided into three blocks as shown in Fig. \ref{fig:matrixH}, corresponding to the three parts of the codeword $\mathbf c$.
Bob must use the whole matrix to decode for both the secret and random messages (since he does not know in advance any of them), although in the end he is interested only in $\mathcal M$.
Eve is in the same condition, although she receives the signal through a different channel.
Frank, instead, has perfect knowledge of $\mathcal M$, and only needs to decode for $\mathcal R$.
Therefore, he can precompute $\mathbf A \cdot {\mathcal M}^T = \mathbf s$, were $^T$ denotes transposition.
Then, he can use $\mathbf s$ as a syndrome vector and focus on the reduced parity-check system:

\[
\left[\mathbf B | \mathbf C \right] \cdot \left[ \mathcal R | \mathcal P \right]^T = \mathbf H' \cdot  \mathbf c'^T = \mathbf s.
\]

Obviously, decoding for a vector having an all-zero syndrome or a different syndrome is equivalent, due to the code linearity.
Hence, Frank performs decoding through the \ac{LDPC} code defined by $\mathbf H'$, having rate $R_F = k_r / (k_r + r)$.
The code rate for Bob instead coincides with the overall code rate, i.e., $R_B = k/n$.
It follows that $R_F = \frac{R_B - R_s}{1 - R_s}$.
In the setting we consider, it is important that both Bob's and Frank's codes are optimized.
In fact, an optimized code for Bob allows to approach the channel capacity, which is the ultimate limit for the reliability target.
An optimized code for Frank instead serves to achieve the desired $\eta$ with the smallest possible \ac{SNR}.
Since Frank's \ac{SNR} is the same as Eve's, this reduces Eve's channel capacity $C_E$.

\section{Code optimization}
\label{sec:CodeOptimization}

We propose an optimization strategy for Bob's and Frank's codes based on the density evolution algorithm, which is commonly 
used to optimize a single code, with some modifications in order to consider the joint optimization target.
In Section \ref{sec:SingleOptimization} we briefly recall the steps of the single code optimization and then in Section \ref{sec:JointOptimization} 
we describe our strategy for the joint code optimization.
In this work, as in \cite{Chung 2001}, we use the density evolution algorithm with Gaussian approximation of the decoder messages.

\subsection{Single optimization}
\label{sec:SingleOptimization}

The density evolution algorithm is well-known in the literature; therefore, for the sake of brevity, we report here only the main equations of \cite{Chung 2001}, as they are used in the proposed joint code optimization. 

Given $\rho(x)$, $R_c$ and $d_{v}$, the optimization of $\lambda(x)$ of a single code is possible by applying the following constraints:

\begin{itemize}
\item [$C_1$] - Rate constraint:
\begin{equation}
\sum_{i=2}^{d_{v}}\frac{\lambda_i}{i} =\frac{1}{1-R_c}\sum_{i=2}^{d_{c}}\frac{\rho_i}{i}.
\label{eq:RateConstraints}
\end{equation} 
\\
\item [$C_2$] - Proportion distribution constraint:
\begin{equation}
\sum_{i=2}^{d_{v}}\lambda_i=1.
\label{eq:ProportionConstraints}
\end{equation} 
\\
\item [$C_3$] - Convergence constraint (from \cite[Eq. (16)]{Chung 2001}):
\begin{equation}
r > h(s, r), \    \  \forall r\in(0,\phi(s))
\label{eq:ConvergenceConstraints}
\end{equation} 
where $s = \frac{2}{\sigma^2}$, $\sigma^2$ being the noise variance, and $\phi(\cdot)$ will be defined in \eqref{eq:phi}.
For $0<s<\infty$ and $0<r\leq1$, we define $h(s, r)$ in (\ref{eq:ConvergenceConstraints}) as follows:

\begin{equation}
h(s, r)=\sum_{i=2}^{d_{v}}\lambda_ih_i(s,r)
\label{eq:h}
\end{equation} 
where
\begin{equation}
h_i(s, r)= \phi\left(s+\left(i-1\right)\sum_{j=2}^{d_{c}}\rho_j\phi^{-1}\left(1-\left(1-r\right)^{j-1}\right)\right).
\label{eq:h_i(s,r)}
\end{equation}

In (\ref{eq:ConvergenceConstraints}) and (\ref{eq:h_i(s,r)}),
\begin{equation}
\phi(x) =
\begin{cases}
1-\frac{1}{\sqrt{4\pi x}}\int_{-\infty}^{+\infty} \tanh\frac{u}{2}e^{-\frac{\left(u-x\right)^2}{4x}}du,      & \mbox{if } x>0 \\
1,                & \mbox{if }  x=0.
\label{eq:phi}
\end{cases}
\end{equation} 

\end{itemize}

\begin{itemize}
\item [ ] Condition \eqref{eq:ConvergenceConstraints} is equivalent to impose that $r_l(s) \rightarrow 0$ for $l \rightarrow \infty$ \cite{Chung 2001}, 
with $r_l=h(s, r_{l-1})$ and $r_0=\phi(s)$.
\\
\item [$C_4$] - Stability condition:
\begin{equation}
\lambda_2 < \frac{e^{\frac{1}{2 \sigma^2}}}{\sum_{j=2}^{d_{c}}\rho_j(j-1)}.
\label{eq:StabilityCondition}
\end{equation}
\end{itemize}
In the single code optimization, the code threshold $s^*$ is defined as the minimum $s$ for which the constraints $[C_1-C_4]$ are satisfied. 
From the definition of $s$, it is evident that $s^*$ corresponds to the maximum noise variance $\sigma^2$ for which the constraints are verified.


\subsection{Joint optimization}
\label{sec:JointOptimization}

In order to perform the joint optimization of Bob's and Frank's codes, we must impose that Frank's code is somehow \textit{contained} in Bob's code (in other terms, that Frank's parity-check matrix is a sub-matrix of Bob's parity-check matrix).
Therefore, in addition to the constraints in Section \ref{sec:SingleOptimization}, we need another condition.
To obtain this further constraint, we introduce the polynomial $\tilde{\lambda}(x)$ which corresponds to the node perspective of $\lambda(x)$. In $\tilde{\lambda}(x)$, the fraction of nodes of degree $i$ can be derived from $\lambda(x)$ through the following formula:
\begin{equation}
\tilde{\lambda}_i=\frac{{\lambda}_i / i}{\sum_{k=2}^{d_{v}}\lambda_k / k}.
\label{eq:DegreeDistributionNode}
\end{equation}
In order to obtain the check node degree distributions from the node perspective $\tilde{\rho}(x)$, a similar formula can be applied to the check nodes degree distributions from the edge perspective. This can be easily achieved by replacing in (\ref{eq:DegreeDistributionNode}) $\lambda(x)$ with $\rho(x)$, $\tilde{\lambda}(x)$ with $\tilde{\rho}(x)$ and $d_v$ with $d_c$.

Since Bob's parity-check matrix contains Frank's parity-check matrix, the number of variable nodes in Bob's Tanner graph having some fixed degree must be greater than or equal to that of variable nodes in Frank's Tanner graph having the same degree.
Hence, we must take into account the following further constraint:

\begin{itemize}
\item [$C_5$] - Joint optimization constraint:
\begin{equation}
\tilde{\lambda}_{B,i} \geq \tilde{\lambda}_{F,i}, \ \ \forall i \in \left[2, 3, 4, \ldots, d_v^{(F)} \right],
\label{eq:JointOptimization}
\end{equation}
\end{itemize}
where $\tilde{\lambda}_B(x)$ and $\tilde{\lambda}_F(x)$ are Bob's and Frank's variable node degree distributions from the node perspective, respectively, and $d_v^{(F)}$ is Frank's maximum variable node degree.
$C_5$ adds to $[C_1-C_4]$ and the optimum $\lambda_B(x)$ must satisfy all these constraints.

In the joint optimization algorithm, we define the convergence threshold as the maximum of $c=\sigma^2_B+\sigma^2_F$, denoted by $c^*$, for which the constraints $[C_1-C_5]$ are satisfied.
In the expression of $c$, $\sigma^2_B$ and $\sigma^2_F$ are Bob's and Frank's noise variances, respectively.
It should be noted that this procedure differs from optimizing the two codes separately.
In fact, in principle, we could first optimize Frank's code, and then try to optimize Bob's code by taking account the degree distributions obtained for Frank and the constraint $C_5$.
This, however, could impose too strong constraints on Bob's code degree distribution, thus preventing to find a good solution for him, too.
In fact, some solutions may exist for which neither Bob's nor Frank's degree distributions are individually optimal, but their joint performance is optimal. 

As in \cite{Chung 2001}, in order to design the check node degree distribution,  we adopt a concentrated distribution (i.e., with only two degrees, concentrated around the mean).
It is widely recognized that this solution, though very simple, is able to achieve very good performance.
Hence, for each pair $(\lambda_F(x), \lambda_B(x))$, we obtain the pair $(\tilde{\rho}_F(x), \tilde{\rho}_B(x))$ by using the following formula, valid for both Bob and Frank:

\begin{equation}
\label{eq:notation1}
\tilde{\rho}(x) = a x^{\left\lfloor c_m \right\rfloor} + b x^{\left\lceil c_m \right\rceil},
\end{equation}
where $c_m = \frac{E}{r} = \frac{\sum_{j} \tilde{\lambda}_j \cdot j}{(1-R_c)}$ and $E$ is the total number of edges in the Tanner graph.
The values $a$ and $b$ are computed as
\begin{equation}
\label{eq:notation2}
a = \lceil c_m \rceil - c_m, \ \ \ \ b = c_m - \lfloor c_m \rfloor.
\end{equation}
In \eqref{eq:notation1} and \eqref{eq:notation2}, $\lceil c_m \rceil$ and $\lfloor c_m \rfloor$ represent the ceiling and floor value of $c_m$, respectively.

\section{Numerical results}
\label{sec:NumericalResults}

In order to provide some practical examples, we use the procedure described in Section \ref{sec:JointOptimization} to design several codes with $d_{v}^{(B)}=d_{v}^{(F)}=50$.
We consider code rates $R_c = R_B = 0.35, 0.5, 0.75$ and several values of $R_s < R_B$.
The degree distributions obtained through the joint optimization procedure are reported in Table \ref{tab:DegreesDistributions}.
Concerning the choice of the degrees of $x$ allowed in the two polynomials, the only constraints we impose are that they must not overcome the maximum values $d_{v}^{(B)}$ and $d_{v}^{(F)}$, and that the number of nodes of degree $2$ must be such that the stability condition \eqref{eq:StabilityCondition} is met by both codes.

\begin{table*}[!t] 
\renewcommand{\arraystretch}{1.1}
\caption{Degrees distribution pairs obtained with the technique described in Section \ref{sec:JointOptimization} for several values of $R_s$ and $R_B$.}
\label{tab:DegreesDistributions}
\centering
\begin{tabular}{|l|c|c|c|c|c|c|c|c|c|c|c|c|c|c|}
\hline
\multicolumn{1}{|l|}{$R_s$} & \multicolumn{2}{|c|}{0.33} &\multicolumn{2}{c}{0.4} & \multicolumn{2}{|c|}{0.45} & \multicolumn{2}{c}{0.5}  & \multicolumn{2}{|c|}{0.6} & \multicolumn{2}{c}{0.7} & \multicolumn{2}{|c|}{0.725}\\
\hline
\multicolumn{1}{|l|}{$R_B$} & \multicolumn{2}{|c|}{0.35} & \multicolumn{2}{c}{0.5} & \multicolumn{2}{|c|}{0.5} & \multicolumn{2}{c}{0.75}  & \multicolumn{2}{|c|}{0.75} & \multicolumn{2}{c}{0.75} & \multicolumn{2}{|c|}{0.75}\\
\hline
$i$&$\lambda_{F,i}$&$\lambda_{B,i}$&$\lambda_{F,i}$&$\lambda_{B,i}$&$\lambda_{F,i}$&$\lambda_{B,i}$&$\lambda_{F,i}$&$\lambda_{B,i}$&$\lambda_{F,i}$&$\lambda_{B,i}$&$\lambda_{F,i}$&$\lambda_{B,i}$&$\lambda_{F,i}$&$\lambda_{B,i}$\\
\hline
\hline
 2 & 0.6677 & 0.1858 & 0.4208 & 0.2259 & 0.6181 & 0.2070 & 0.2187 & 0.1588 & 0.2066 & 0.1382 & 0.4257 & 0.1712 & 0.6181 & 0.1300  \\
\hline
 3 & 0.2279 & 0.2291 &  0.1656 & 0.1701 & 0.2117 & 0.2123 & 0.1826 & 0.1851 & 0.1436 & 0.1549 & 0.1763 & 0.1787 & 0.2117 & 0.2128  \\
\hline 
 4 &    -   &    -   &  0.1192 & 0.1195 &    -   &   -    &    -   &    -   & 0.0280 & 0.0278 & 0.1014 & 0.1029 &    -   &    -    \\ 
\hline
 5 &    -   &    -   &    -   &    -   & 0.1445  & 0.1471 & 0.0497 & 0.0449 & 0.0123 & 0.0112 &    -   &    -   & 0.1445 & 0.1786 \\
\hline
 6 & 0.0267 & 0.0252 &     -  &    -   & 0.0246  & 0.0254 & 0.0365 & 0.0378 & 0.0248 & 0.0267 &    -   &    -   & 0.0246 & 0.0354 \\
\hline 
 7 & 0.0767 & 0.0751 &    -   &   -    &    -   &    -   &  0.0309 & 0.0317 & 0.0999 & 0.1054 &    -   &    -   &   -    &    -   \\
\hline
 8 &    -   &    -   &  0.0057 & 0.0061 &    -   &    -   & 0.1662 & 0.1683 &    -   &    -   &    -   &    -   &    -   &    -   \\
\hline 
 9 &    -   &    -   &     -   &    -   &    -   &    -   &    -   &    -   & 0.0539 & 0.0574 & 0.1321 & 0.1410 &    -   &    -   \\
\hline 
 10 &    -   &    -   & 0.2877 & 0.2907 &    -   &    -    &    -   &    -   & 0.0413 & 0.0409 & 0.1635 & 0.1639 &    -   &    -   \\
\hline
 11 &    -   & 0.0249 &   -   &    -   &    -   &    -   &    -   &    -   & 0.0144 & 0.0175 &    -   &    -   &    -   &    -   \\
\hline
 12 &    -   & 0.1792 &    -   &    -   &    -   &    -   &    -   &    -   & 0.0126 & 0.0119 &    -   &    -   &    -   &    -   \\
\hline
 13 &    -   &    -   &    -   &    -   &    -   &    -   &    -   &    -   &    -   &    -   &    -   &    -   &    -   & 0.0359 \\
\hline   
 14 &    -   &    -   &    -   &    -   &    -   & 0.0184 &    -   &    -   &    -   &    -   &    -   &    -   &    -   & 0.0625 \\
\hline
 15 &    -   &    -   &    -   &    -   &    -   & 0.2779 &    -   &    -   &    -   &    -   &    -   &    -   &    -   & 0.1561 \\
\hline
 19 &    -   &    -   &    -   &   -    &    -   &    -   &   -    &    -   & 0.0637 & 0.0713 &    -   &   -    &    -   &    -   \\
\hline
 20 &    -   &    -   &    -   &   -    &    -   &    -   & 0.0154 & 0.0124 & 0.0050 & 0.0190 &    -   &    -   &    -   & 0.0031 \\
\hline
 21 &    -   &    -   &    -   & 0.0096 &    -   &   -    & 0.0747 & 0.0954 &    -   &    -   &    -   &    -   &    -   & 0.0103  \\
\hline
 22 &    -   &    -   &    -   &   -    &    -   &    -   & 0.0666 & 0.0659 &    -   &    -   &    -   &    -   &    -   & 0.0014 \\
\hline
 23 &    -   &    -   &    -   &    -   &    -   & 0.1109 & 0.0568 & 0.0549 &    -   &    -   &    -   &    -   &    -   &    -   \\
\hline
 24 &    -   &    -   &    -   &   -    &    -   &    -   &    -   &    -   &    -   &    -   &    -   & 0.0307 &    -   &    -   \\
\hline
 25 &    -   &    -   &    -   & 0.0697 &    -   &   -    & 0.1007 & 0.1016 &    -   &    -   &    -   & 0.2106 &    -   &    -   \\
\hline
 26 &    -   &    -   &    -   & 0.1074 &    -   &   -    &    -   &    -   &    -   &    -   &    -   &    -   &    -   &    -   \\
\hline
 32 &    -   &    -   &    -   &    -   &    -   &    -   &    -   &    -   &    -   &    -   &    -   &    -   &    -   & 0.1727 \\
\hline
 34 &    -   & 0.0203 &    -   &    -   &    -   &    -   &    -   &    -   &    -   &    -   &    -   &    -   &    -   &    -   \\
\hline
 36 &    -   & 0.0844 &    -   &    -   &    -   &    -   &    -   &    -   &    -   &    -   &    -   &    -   &    -   &    -   \\
\hline
 38 &    -   & 0.0716&    -   &    -   &    -   &    -   &    -   &    -   &    -   &    -   &    -   &    -   &    -   &    -   \\
\hline
 39 &    -   & 0.0652 &    -   &    -   &    -   &    -   &    -   &    -   & 0.2929 & 0.2946 &    -   &    -   &    -   &    -   \\
\hline
 40 &    -   & 0.0382 &    -   &    -   &    -   &    -   &    -   &    -   &    -   &    -   &    -   &    -   &    -   &    -   \\
\hline 
 50 & 0.0010 & 0.0010 & 0.0010 & 0.0010 & 0.0011 & 0.0010 & 0.0012 & 0.0432 & 0.0010 & 0.0232 & 0.0010 & 0.0010 & 0.0011 & 0.0012 \\
\hline
\end{tabular}
\end{table*}

To provide some examples of finite length codes, we consider \ac{LDPC} codes with length $n_1=10000$ and $n_2=50000$; Frank's code length is then obtained from these values by considering the submatrix $\mathbf H'$.
Once having defined the degree distributions, the parity-check matrices are designed through the \ac{PEG} algorithm \cite{Hu2001PEG}.
The numerical results are obtained by considering, for all coding schemes, \ac{BPSK} modulation over the \ac{AWGN} channel.
When considering finite length codes, through numerical simulations we are able to determine the values of the \ac{SNR} per bit $(E_b/N_0)$ that ensure a given \ac{CER}. These values are reported in Table \ref{tab:Performance}, for both Bob and Frank, assuming $\ac{CER} = 10^{-2}$ and several values of $R_s$.
In the table, the values of $\left.\frac{E_b}{N_0}\right|_{th}$ identify the codes convergence thresholds obtained through density evolution.
These values represent the ultimate performance bounds achievable in asymptotic conditions (i.e., infinite code length).
The values of $\left.\frac{E_b}{N_0}\right|_{n_1}$ and $\left.\frac{E_b}{N_0}\right|_{n_2}$ instead represent the \ac{SNR} working points, estimated through simulations, for the practical codes with lengths $n_1$ and $n_2$, respectively.
We observe from Table \ref{tab:Performance} that, for Bob's code, the finite length performance approaches the asymptotic threshold as the code rate increases.
Indeed, for $R_B = 0.75$ and code length equal to $n_1$ and $n_2$, the gap between the asymptotic threshold and the finite length codes performance is about $0.4$ dB and $0.2$ dB, respectively. 

\begin{table}[!t] 
\renewcommand{\arraystretch}{1.1}
\caption{\ac{SNR} working points of the considered coding schemes for several values of $R_s$ and $R_B$; the values of $\frac{E_b}{N_0}$ are in dB.}
\label{tab:Performance}
\centering
\begin{tabular}{|@{\hspace{1mm}}c@{\hspace{1mm}}|@{\hspace{1mm}}c@{\hspace{1mm}}|@{\hspace{1mm}}c@{\hspace{1mm}}|@{\hspace{1mm}}c@{\hspace{1mm}}|@{\hspace{1mm}}c@{\hspace{1mm}}|@{\hspace{1mm}}c@{\hspace{1mm}}|@{\hspace{1mm}}c@{\hspace{1mm}}|@{\hspace{1mm}}c@{\hspace{1mm}}|}
\hline
$R_s$&$R_B$&$\left.\frac{E_b}{N_0}\right|^{B}_{th}$&$\left.\frac{E_b}{N_0}\right|^{F}_{th}$&$\left.\frac{E_b}{N_0}\right|^{B}_{n_1}$&$\left.\frac{E_b}{N_0}\right|^{F}_{n_1}$&$\left.\frac{E_b}{N_0}\right|^{B}_{n_2}$&$\left.\frac{E_b}{N_0}\right|^{F}_{n_2}$\\
\hline
0.33  & 0.35 & -0.14 & -1.52 & 1.10 & 3.82 & 0.72 & 3.18 \\ 
\hline
 0.4  & 0.5 & 0.41 & -0.52 & 1.00 & 0.76 & 0.78 &  0.44 \\ 
\hline
 0.45 & 0.5 & 0.42 & -0.69 & 1.12 & 1.22 & 0.82 & 0.98 \\
\hline
 0.5 & 0.75 & 1.73 & 0.38  & 2.14 & 1.17 & 1.94 &  0.84\\
\hline
 0.6 & 0.75 & 1.72 & -0.14 & 2.12 & 0.98 & 1.97 &  0.63\\
\hline
 0.7 & 0.75 & 1.75 & -0.52 & 2.13 & 0.91 & 1.92 & 0.60 \\
\hline
 0.725 & 0.75 & 1.75 & -0.69 & 2.18 & 2.11 & 1.96 &  1.59 \\
\hline
\end{tabular}
\end{table}

\begin{figure}[!t]
\begin{centering}
\includegraphics[width=80mm,keepaspectratio]{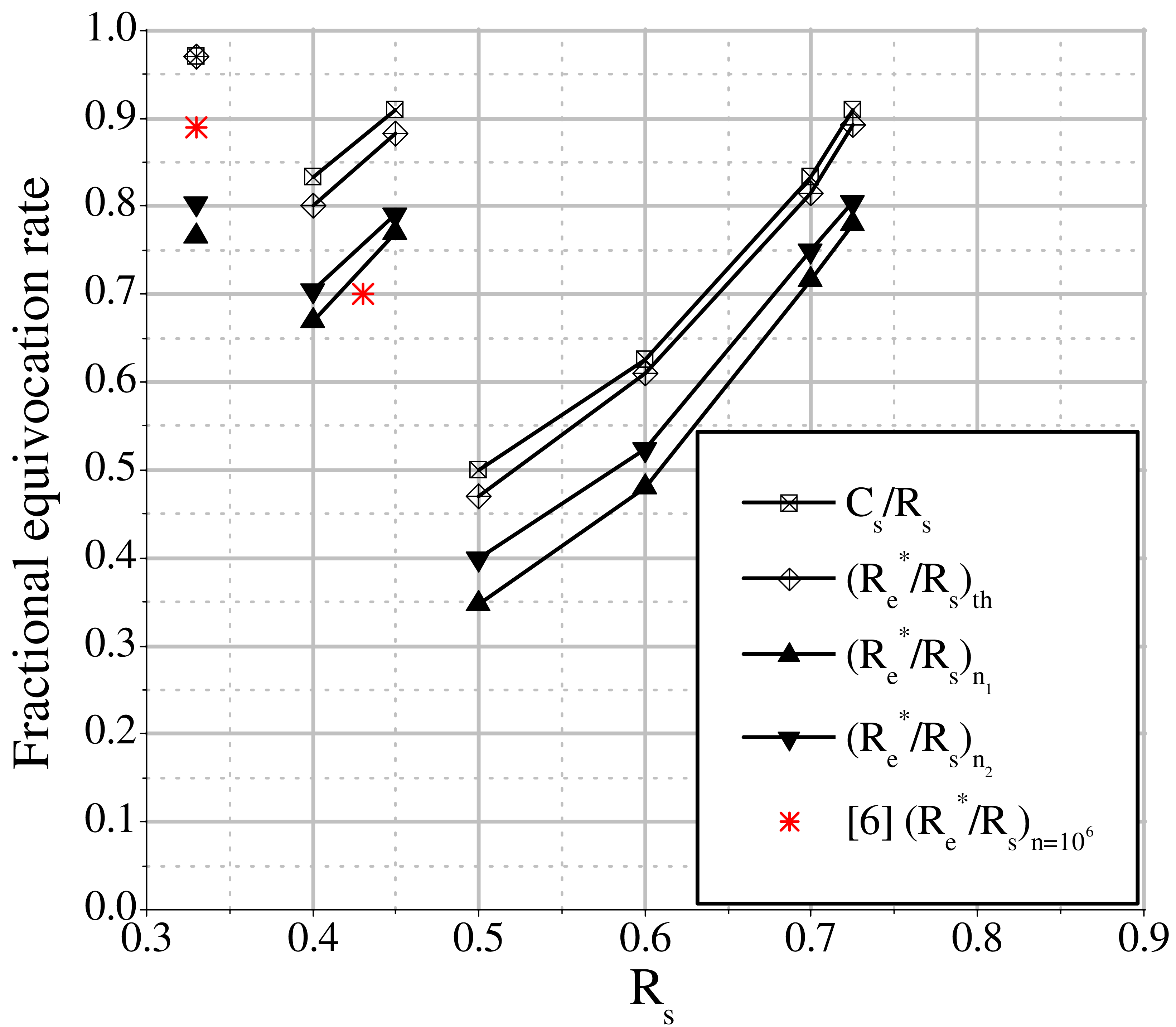}
\caption{Comparison between $\frac{C_s}{R_s}$, $\frac{R_e^*}{R_s}$ calculated through the asymptotic threshold values, $\frac{R_e^*}{R_s}$ for code length $n_1$, and $\frac{R_e^*}{R_s}$ for code length $n_2$, as a function of ${R_s}$.}
\label{fig:ReRs}
\par\end{centering}
\end{figure}

As a security metric we use the lower bound $R_e^*$ on the equivocation rate, computed according to \eqref{eq:ReBound} and the values in Table \ref{tab:Performance}.
The secrecy capacity $C_s$, that represents the ultimate limit achievable by the equivocation rate, is also computed for the cases of interest, and used as a benchmark.
We compute $C_s$ under the hypothesis of ideal coding, i.e., that Bob's and Frank's code rates coincide with the respective channel capacities.
Since Frank's and Eve's channels coincide, it follows that $C_s = R_B - R_F = R_s \frac{1-R_B}{1-R_s}$.
In order to assess if practical codes can approach the perfect secrecy condition \eqref{eq:PerfectSecrecy}, we then compute the fractional lower bound on the equivocation rate $\widetilde{R_e^*} = R_e^*/R_s$
both in asymptotic conditions and in the finite code length regime, and compare its values with the fractional secrecy capacity $\widetilde{C_s} = C_s/R_s = \frac{1-R_B}{1-R_s}$.
The values so obtained are reported in Fig. \ref{fig:ReRs}, for the same values of $R_s$ considered in Tables \ref{tab:DegreesDistributions} and \ref{tab:Performance}.
As an example, for the considered code parameters and $R_s = 0.725$, we find that in asymptotic conditions the designed codes approach the secrecy capacity and the perfect secrecy condition.
Notably, even using relatively short codes, with $10000$-bit codewords, the fractional equivocation rate is close to $0.8$.
For the sake of comparison, we consider some results reported in \cite{WongWong2011a} for the scheme based on punctured \ac{LDPC} codes.
The corresponding points are marked with an asterisk in Fig. \ref{fig:ReRs}.
Those results consider codes with length $n=10^6$, at which the performance of \ac{LDPC} codes usually approaches the density evolution threshold.
However, the asymptotic performance achieved by the degree distributions found through the proposed approach exhibits some gain at the same secret message rates.
Furthermore, for $R_s = 0.43$, even our schemes with $n=10000$ and $n=50000$ outperform that proposed in \cite{WongWong2011a} with $n=10^6$.

From Fig. \ref{fig:ReRs} it results that the best performance in terms of Eve's equivocation rate is achieved when the secret message rate approaches the code rate.
This could seem counterintuitive, since suggests to use few random bits to confuse the eavesdropper.
However, in this condition $R_F$ is small and Frank is able to reach the desired performance at low \ac{SNR}.
The latter coincides with Eve's channel \ac{SNR}, therefore Eve's equivocation rate is large.
On the other hand, imposing that Eve's channel has a too low \ac{SNR} is not realistic, therefore some randomness shall always be used in order to relax the constraints on Eve's channel quality.

\section{Conclusion}
\label{sec:Conclusion}

We have studied the performance of practical \ac{LDPC} coded transmissions over the Gaussian wiretap channel.
By using suitable reliability and security metrics, we have computed performance bounds in the asymptotic regime
and assessed the achievable performance under the hypothesis of finite codeword lengths.
We have also proposed an optimization approach to design good codes for this context.
Our results show that these codes are able to approach the ultimate performance limits even with relatively small block lengths.

\newcommand{\BIBdecl}{\setlength{\itemsep}{0.005\baselineskip}}
\bibliographystyle{IEEEtran}
\bibliography{Archive}

\end{document}